\documentclass{PoS}

\usepackage{amsmath,amssymb,siunitx,booktabs}
\usepackage{graphicx,subcaption}
\newcommand{\Li}{\operatorname{Li}}
\allowdisplaybreaks

\title{The effect of improved high-energy muon cross-sections}

\ShortTitle{Muon cross-sections}

\author{\speaker{Jan Soedingrekso}, Alexander Sandrock and Wolfgang Rhode\\
        Technische Universit\"at Dortmund, Germany\\
				E-mail: \email{jan.soedingrekso@tu-dortmund.de},
				\email{alexander.sandrock@tu-dortmund.de}}

\abstract{
	A precise simulation of muons with energies above a TeV is crucial for
	neutrino telescopes or cosmic ray experiments. To further increase the
	precision of these simulations, improved cross-section calculations are
	needed. At these energies, the interaction probability is dominated by
	bremsstrahlung for large energy losses and electron-positron pair production
	for small energy losses.
	
	Improved analytical calculations for these processes were developed with more
	precise screening functions of the target atom as well as higher order
	corrections reducing the systematic uncertainties below the percent level.
	The new calculations are already implemented in the new version of the lepton
	propagator PROPOSAL, which was designed to be highly performant for the
	propagation through large volumes of media using interpolation tables and to
	do systematic studies with its multiple available cross-section calculations.
	The new calculations of the differential cross section result in a maximum
	deviation of 3 percent to the current standard. Their effects on the muon
	simulation with the resulting flux and energy loss distribution are
	presented.}

\FullConference{36th International Cosmic Ray Conference -ICRC2019-\\
		July 24th - August 1st, 2019\\
		Madison, WI, U.S.A.}

\begin{document}

\section{Improved cross sections}
\subsection{Leading-order bremsstrahlung cross section}
The singly-differential cross section for bremsstrahlung of highly-relativistic
leptons in the field of an atomic nucleus with charge $Z e$ and mass $A m_N$
can be expressed as \cite{BetheHeitler}
\begin{equation}
	d\sigma = 4 \alpha \left(Z r_e \frac{m_e}{\mu}\right)^2 \frac{dv}{v}
	\left[(2 - 2 v + v^2) \Phi_1(Z, A, \delta) - \frac{2}{3} (1 - v)
	\Phi_2(Z, A, \delta) \right],
\end{equation}
where $\mu$ refers to the mass of the lepton, $m_e$ the electron mass,
$\alpha$ is the fine structure constant and $r_e$ the classical electron radius.
$v$ is the fractional energy loss of the lepton in the interaction.
The dependence on the electric field of the nucleus is contained in the
screening functions $\Phi_{1,2}$ which only depend on the nucleus and the
minimum momentum transfer
\begin{equation}
	\delta = \frac{\mu^2 v}{2 E (1 - v)}.
\end{equation}

In the limiting case of complete screening and a point-like nucleus, the
screening functions become independent of $\delta$ and are given by
\begin{equation}
	\Phi_1^0 = \ln \left(B \frac{\mu}{m_e} Z^{-1/3}\right),\quad
	\Phi_2^0 = \Phi_1^0 - \frac{1}{6}.
\end{equation}
Here, $B$ is the radiation logarithm. In the Thomas-Fermi model, $B = 183$,
while in the more exact Hartree-Fock model $B$ depends on the nucleus. For the
simulations, we use the Hartree-Fock radiation logarithms from
\cite{RadiationLogarithm}. In the absence of screening, the screening functions
coincide for a point-like nucleus
\begin{equation}
	\Phi_1^0 = \Phi_2^0 = \ln\frac{\mu}{\delta} - \frac{1}{2}.
\end{equation}
In the approximation $\Phi_1 \approx \Phi_2$, an analytical interpolation was
found by \cite{PetrukhinShestakov}
\begin{equation}
	\Phi = \ln \frac{(\mu/m_e) B Z^{-1/3}}{1 + (\delta/m_e) \sqrt{e} B
	Z^{-1/3}}
\end{equation}
which describes the intermediate behavior between complete screening and no
screening for a point-like nucleus with high accuracy for medium and heavy
nuclei.

The correction $\Delta_{1,2}$ for a nuclear form factor due to the finite size
of the nucleus on the screening functions $\Phi_{1,2}^0$ is independent of
screening, as very different regimes of transferred momenta to the nucleus are
concerned, such that $\Phi_i = \Phi_i^0 - \Delta_i$.

Numerical results for the nuclear correction, calculated using a Fermi form
factor with nuclear sizes according to \cite{Elton}, can be fitted with good
accuracy using the expression
\begin{equation}
	\Delta_1 = \ln \frac{\mu}{q_c} + \frac{\rho}{2} \ln \frac{\rho + 1}
	{\rho - 1},\quad
	\Delta_2 = \ln \frac{\mu}{q_c} + \frac{3 \rho - \rho^3}{4}
	\ln \frac{\rho + 1}{\rho - 1} + \frac{2 \mu^2}{q_c^2}
\end{equation}
with $\rho = \sqrt{1 + 4 \mu^2/q_c^2}$, which follows from a step function for
the nuclear form factor \cite{Bugaev,ABB2}. Fitting to numerical results leads
to $q_c = m_\mu e/D_n$ with $D_n = \num{1.54} A^{\num{0.27}}$ (cf. \cite{KKP}).

We apply the interpolation analogous to \cite{PetrukhinShestakov}
separately to $\Phi_{1,2}$ without assuming the screening functions to be equal.
Adding the contributions due to the inelastic nuclear form factor \cite{ABB2}
and the contribution from atomic electrons \cite{KKP-atomic}, we obtain the
improved leading-order cross section for bremsstrahlung as
\begin{equation}
	\begin{split}
		\frac{d\sigma}{dv} &= 4 Z^2 \alpha \left(r_e \frac{m_e}{\mu}\right)^2
		\frac{1}{v} \left\{\left[(2 - 2 v + v^2) \Phi_1(\delta)
		- \frac{2}{3} (1 - v) \Phi_2(\delta)\right] + \frac{1}{Z}
		s_\text{atomic}(v, \delta)\right\},\\
		\Phi_1(\delta) &= \ln \frac{\frac{\mu}{m_e} B Z^{-1/3}}{1 + B Z^{-1/3}
		\sqrt{e} \delta/m_e} - \Delta_1 \left(1 - \frac{1}{Z}\right),\\
		\Phi_2(\delta) &= \ln \frac{\frac{\mu}{m_e} B Z^{-1/3}}{1 + B Z^{-1/3}
		\sqrt[3]{e} \delta/m_e} - \Delta_2 \left(1 - \frac{1}{Z}\right),\\
		s_\text{atomic}(v, \delta) &= \left[\frac{4}{3} (1 - v) + v^2\right]
		\left[\ln \frac{\mu/\delta}{\mu \delta/m_e^2  + \sqrt{e}} - \ln \left(1 +
		\frac{m_e}{\delta B' Z^{-2/3} \sqrt{e}}\right) \right].
	\end{split}
\end{equation}

\subsection{Leading-order pair production cross section}
The processes of pair production and bremsstrahlung are intimately connected.
The discussion of the screening functions for bremsstrahlung can be analogously
applied to the pair production process. However, an additional integration is
carried out to obtain the doubly differential cross section $d^2\sigma/(dv\,
d\rho)$ with $v = (E_+ + E_-)/E$, $\rho = (E_+ - E_-)/(E_+ + E_-)$ designating
the electron (positron) energy by $E_\pm$. The pair production cross section
can be written as (cf. \cite{KokoulinPetrukhin1,KokoulinPetrukhin2})
\begin{equation}
	\frac{d^2\sigma}{dv\,d\rho} = \frac{2}{3\pi} (Z \alpha r_e)^2 \frac{1 - v}{v}
	\left[C_e L_e + \frac{m_e^2}{\mu^2} C_\mu L_\mu\right]
\end{equation}
with
\begin{align}
	C_e &= [(2 + \rho^2) (1 + \beta) + \xi (3 + \rho^2)] \ln\left(1 + \frac{1}
	{\xi}\right) + \frac{1 - \rho^2 - \beta}{1 + \xi} - (3 + \rho^2),\\
	C_\mu &= \left[(1 + \rho^2) \left(1 + \frac{3}{2} \beta \right)
	- \frac{(1 + 2 \beta) (1 - \rho^2)}{\xi}\right] \ln(1 + \xi) + \frac{\xi
	(1 - \rho^2 - \beta)}{1 + \xi} + (1 + 2 \beta) (1 - \rho^2),\\
	\beta &= \frac{v^2}{2 (1 - v)}, \quad \xi = \left(\frac{\mu v}{2 m_e}\right)^2
	\frac{1 - \rho^2}{1 - v}.
\end{align}
The functions $L_{e,\mu}$ correspond to the screening functions integrated over
the momentum transfer to the electron-positron-pair. This additional integration
leads to terms not contained in the main logarithm which can be expressed in the
limiting cases no screening and complete screening for a point-like nucleus by
\begin{align}
	\Phi_e^N &= C_e \ln \frac{E v (1 - \rho^2)}{2 m_e \sqrt{e} \sqrt{1 + \xi}} -
	\frac{1}{2} |\Delta_e^N|,\\
	\Phi_e^S &= C_e \ln (B Z^{-1/3} \sqrt{1 + \xi}) + \frac{1}{2} \Delta_e^S,
\end{align}
where \cite{Kelner67}
\begin{align}
	|\Delta_e^N| &= [(2 + \rho^2) (1 + \beta) + \xi (3 + \rho^2)] \Li_2 \frac{1}
	{1 + \xi} - (2 + \rho^2) \xi \ln \left(1 + \frac{1}{\xi}\right)
	- \frac{\xi + \rho^2 + \beta}{1 + \xi},\\
	\Delta_e^S &= |\Delta_e^N| - \frac{1}{6} \left\{[(1 - \rho^2) (1 + \beta) +
	\xi (3 + \rho^2)] \ln \left(1 + \frac{1}{1 + \xi}\right) + \frac{1 + 2 \beta
	+ \rho^2 + \xi (3 - \rho^2)}{1 + \xi}\right\}.
\end{align}
Here, $\Li_2(x)$ is the dilogarithm defined by $\Li_2(x) = -\operatorname{Re}
\int_0^x \frac{\ln(1 - t)}{t} dt$.

The difference between $|\Delta_e^N|$ and $\Delta_e^S$ had been neglected in
earlier parametrizations of the pair production cross section. This difference
is the effect of the difference in the screening functions $\Phi_{1,2}$ before
the additional integration. In addition, the special functions in $\Delta_e$
were approximated by an expression consisting only of elementary functions in
\cite{KokoulinPetrukhin1}. Based on the results in \cite{Kelner67}, we have
separated the coefficients of the two screening functions. Incorporating the
calculations for the effect of a nuclear form factor analogous to the
bremsstrahlung case and \cite{KokoulinPetrukhin2} and adding the contribution of
atomic electrons \cite{Kelner-atomic}, we arrive at the following expression for
the pair production cross section
\begin{equation}
\frac{d^2\sigma}{dv\,d\rho} = \frac{2}{3\pi} Z (Z + \zeta) \frac{1 - v}{v}
  \left[\Phi_e + \frac{m_e^2}{m_\mu^2} \Phi_\mu\right],
\end{equation}
where 
\begin{align}
\Phi_e &= C_1^e L_1^e + C_2^e L_e^2,\quad C_1^e = C_e - C_2^e,\\
C_2^e &= [(1 - \rho^2)(1 + \beta) + \xi (3 - \rho^2)] \ln\left(1 + \frac{1}
  {\xi}\right) + 2 \frac{1 - \beta - \rho^2}{1 + \xi} - (3 - \rho^2),\\
L_1^e &= \ln \frac{B Z^{-1/3} \sqrt{1 + \xi}}{X_e + \frac{2 m_e \sqrt{e} B
	Z^{-1/3} (1 + \xi)}{E v (1 - \rho^2)}} - \frac{\Delta_e}{C_e}
  - \frac{1}{2} \ln \left[X_e + \left(\frac{m_e}{m_\mu} D_n\right)^2 (1 + \xi)
  \right]\\
L_2^e &= \ln \frac{B Z^{-1/3} e^{-1/6}\sqrt{1 + \xi}}{X_e + \frac{2 m_e
	e^{1/3} B Z^{-1/3} (1 + \xi)}{E v (1 - \rho^2)}} - \frac{\Delta_e}{C_e}
  - \frac{1}{2} \ln \left[X_e + \left(\frac{m_e}{m_\mu} D_n\right)^2 e^{1/3}
  (1 + \xi) \right],\quad
X_e = \exp\left(-\frac{\Delta_e}{C_e}\right),\\
\Delta_e &= [(2 + \rho^2)(1 + \beta) + \xi (3 + \rho^2)] \Li_2 \frac{1}{1 + \xi}
  - (2 + \rho^2) \xi \ln \left(1 + \frac{1}{\xi}\right) - \frac{\xi + \rho^2
  + \beta}{1 + \xi},
\end{align}
and
\begin{align}
\Phi_\mu &= C_1^\mu L_1^\mu + C_2^\mu L_2^\mu,\quad C_1^\mu =  C_\mu - C_2^\mu,\\
C_2^\mu &= [(1 - \beta)(1 - \rho^2) - \xi (1 + \rho^2)] \frac{\ln (1 + \xi)}
  {\xi} - 2 \frac{1 - \beta - \rho^2}{1 + \xi} + 1 - \beta - (1 + \beta)
  \rho^2,\\
L_1^\mu &= \ln\frac{B \frac{\mu}{m_e} Z^{-1/3}/D_n}{X_\mu + \frac{2 m_e \sqrt{e}
  B Z^{-1/3} (1 + \xi)}{E v (1 - \rho^2)}} - \frac{\Delta_\mu}{C_\mu},\\
L_2^\mu &= \ln\frac{B \frac{\mu}{m_e} Z^{-1/3}/D_n}{X_\mu + \frac{2 m_e e^{1/3}
  B Z^{-1/3} (1 + \xi)}{E v (1 - \rho^2)}} - \frac{\Delta_\mu}{C_\mu},\quad
X_\mu = \exp\left(-\frac{\Delta_\mu}{C_\mu}\right),\\
\begin{split}
\Delta_\mu &= \left[(1 + \rho^2) \left(1 + \frac{3}{2} \beta\right)
  - \frac{1}{\xi} (1 + 2 \beta) (1 - \rho^2)\right] \Li_2\left(\frac{\xi}{1
  + \xi}\right)\\ &+ \left(1 + \frac{3}{2} \beta\right) \frac{1 - \rho^2}{\xi}
  \ln (1 + \xi) + \left[1 - \rho^2 - \frac{\beta}{2} (1 + \rho^2) + 
  \frac{1 - \rho^2}{2 \xi} \beta\right] \frac{\xi}{1 + \xi}
\end{split}
\end{align}
with the abbreviations
\begin{align}
\zeta &= \frac{0.073 \ln \frac{E/\mu}{1 + \gamma_1 Z^{2/3} E/\mu} - 0.26}
  {0.058 \ln \frac{E/\mu}{1 + \gamma_2 Z^{1/3} E/\mu} - 0.14},\\
\gamma_1 &= \num{1.95e-5}, \quad \gamma_2 = \num{5.3e-5} \text{ for } Z \neq 1,
  \\
\gamma_1 &= \num{4.4e-5}, \quad \gamma_2 = \num{4.8e-5} \text{ for } Z = 1.
\end{align}

\subsection{Radiative corrections to bremsstrahlung}
The radiative corrections to the bremsstrahlung cross section have been
calculated based on the Weizs\"{a}cker-Williams method and the radiative
corrections to the Compton cross section \cite{BrownFeynman} and the double
Compton cross section \cite{Skyrme}. In \cite{SandrockKelnerRhode}, some of us
have calculated radiative corrections to the average energy loss. Based on this
calculation we have determined the corrections to the differential cross
section. The cross section factorizes into the screening function and a
universal function $s_\text{rad}(v)$ of the fractional energy loss $v$. The
calculation on the basis of the Weizs\"acker-Williams method cannot distinguish
between the terms proportional to $\Phi_1$ and $\Phi_2$. Since this would be a
small correction of a few percent to the already small radiative corrections,
this does not pose a problem. Numerical calculations can be parametrized by
\begin{equation}
	\begin{split}
		\left.\frac{d\sigma}{dv}\right|_\text{rad} &= Z^2 \alpha^2 \left(r_e
		\frac{m_e} {\mu}\right)^2 \Phi_1(\delta) s_\text{rad}(v),\\
		s_\text{rad}(v) &=\begin{cases}\sum_{n = 0}^2 a_n v^n & v < 0.02,\\
			\sum_{n = 0}^3 b_n v^n & 0.02 \leq v < 0.1,\\
			\sum_{n = 0}^2 c_n v^n + c_3 v \ln v + c_4 \ln(1 - v) + c_5 \ln^2(1 - v)
			& 0.1 \leq v < 0.9,\\
			\sum_{n = 0}^2 d_n v^n + d_3 v \ln v + d_4 \ln(1 - v) + d_5 \ln^2(1 - v)
			& v \geq 0.9,
		\end{cases}
	\end{split}
\end{equation}
where the values of the fit parameters $a_n, b_n, c_n, d_n$ are given in Table~%
\ref{tab:rad_params}.
\begin{table}
\caption{Parameters of the parametrization for the radiative corrections to the
  bremsstrahlung cross section.}
	\begin{center}
		\begin{tabular}{crrrrrr}
			\toprule
			$n$ & 0 & 1 & 2 & 3 & 4 & 5\\
			\midrule
			$a_n$ & $-$0.00349 & 148.84 & $-$987.531\\
			$b_n$ & 0.1642 & 132.573 & $-$585.361 & 1407.77\\
			$c_n$ & $-$2.8922 & $-$19.0156 & 57.698 & $-$63.418 & 14.1166 & 1.84206\\
			$d_n$ & 2134.19 & 581.823 & $-$2708.85 & 4767.05 & 1.52918 & 0.361933\\
			\bottomrule
		\end{tabular}
	\end{center}
	\label{tab:rad_params}
\end{table}

\section{Effect of improved cross sections on muon simulations}
\begin{figure}
    \begin{subfigure}{0.48\textwidth}
        \centering
        \includegraphics[width=\textwidth]{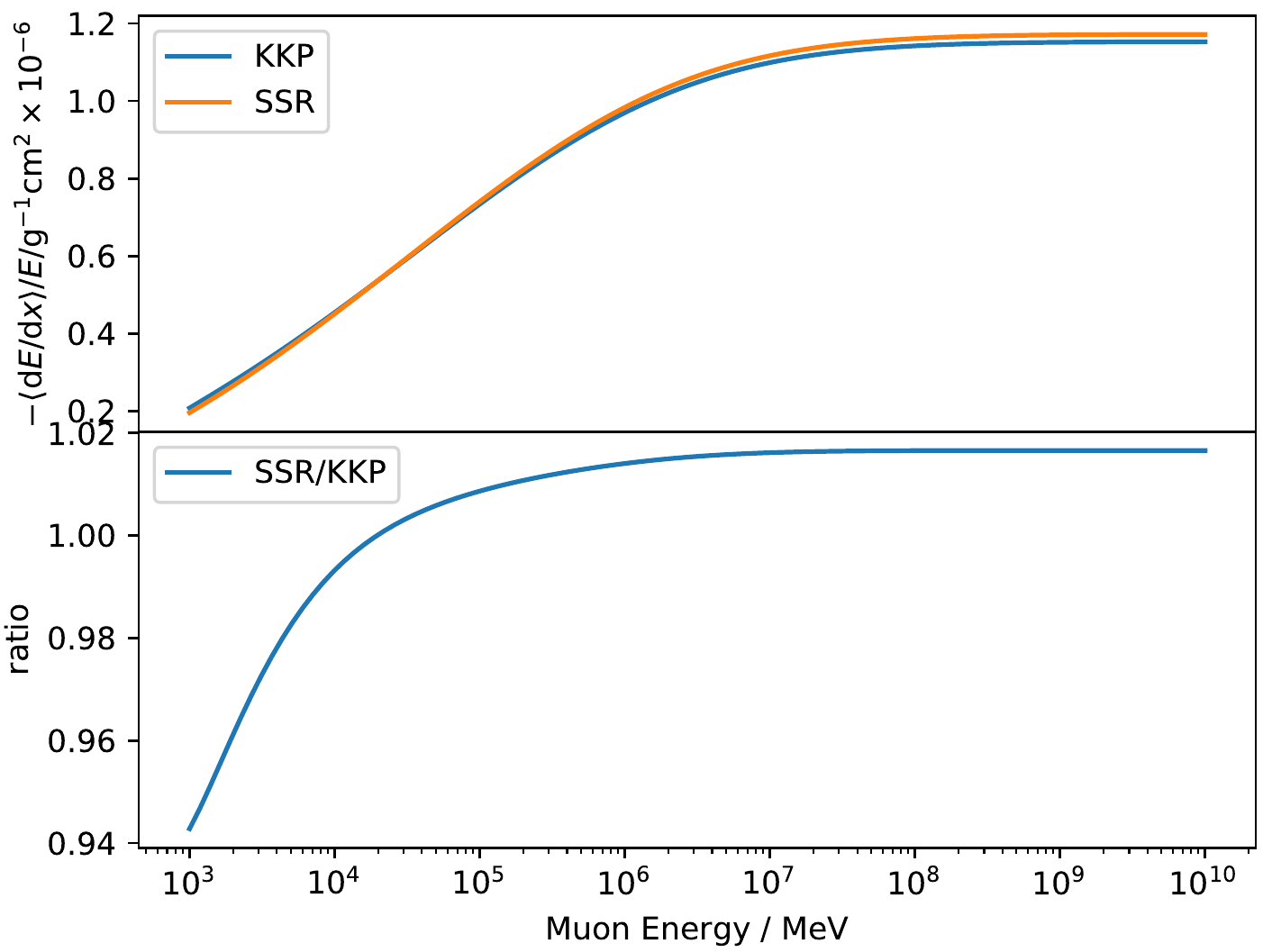}
        \caption{Average energy loss of bremsstrahlung.}
        \label{fig:dedx_brems}
    \end{subfigure}
    \hfill
    \begin{subfigure}{0.48\textwidth}
        \centering
        \includegraphics[width=\textwidth]{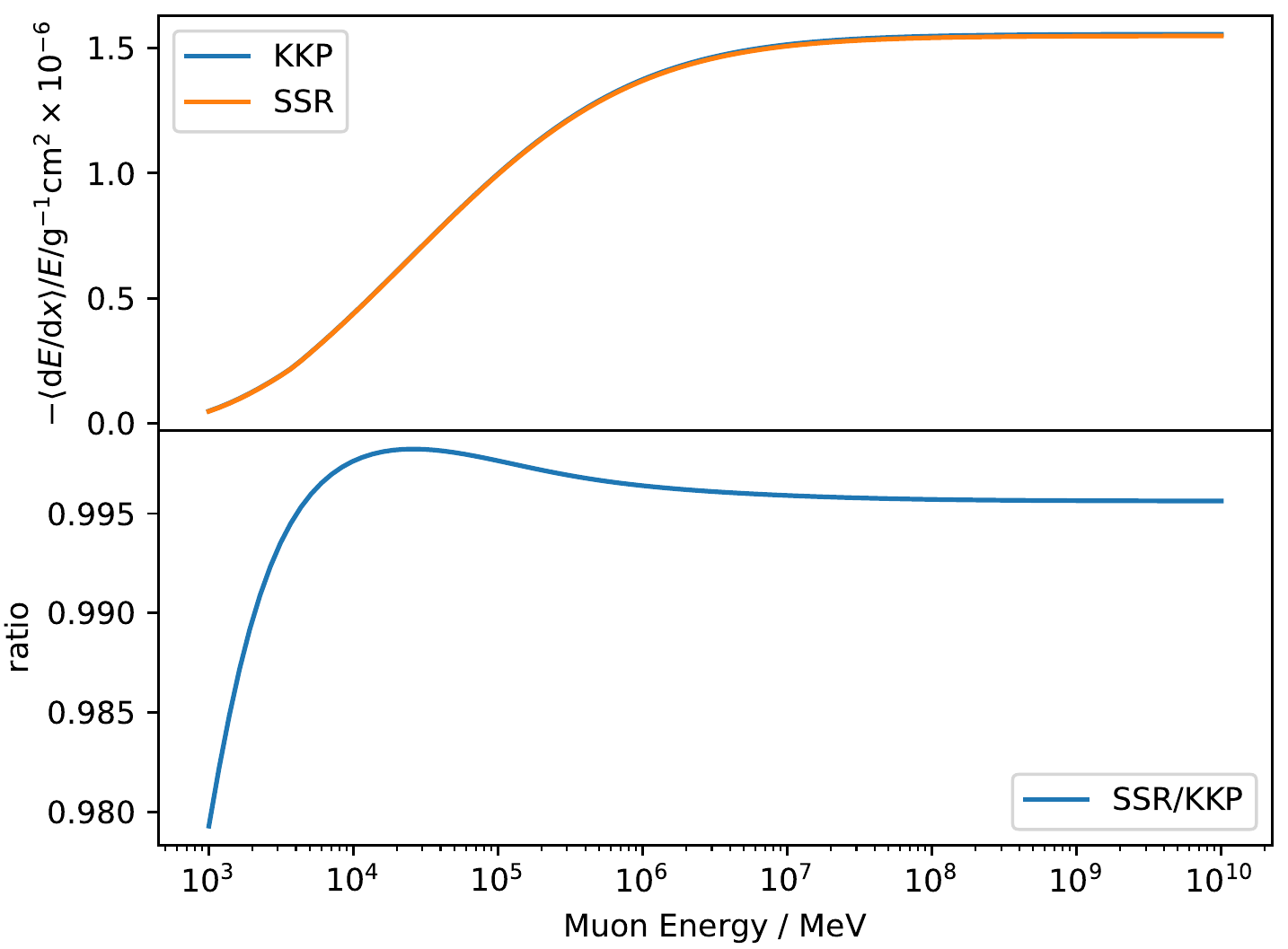}
        \caption{Average energy loss of pair production.}
        \label{fig:dedx_pair}
    \end{subfigure}
		\caption{The average energy loss of the cross sections shown above compared to Kelner et al. \cite{KKP,KKP-atomic,KokoulinPetrukhin1,KokoulinPetrukhin2,Kelner-atomic} for bremsstrahlung \ref{fig:dedx_brems} and pair production \ref{fig:dedx_pair}.}
    \label{fig:dedx_compare}
\end{figure}

The average energy loss of the new bremsstrahlung and pair production cross sections compared to the baseline cross sections is shown in Figure \ref{fig:dedx_compare}.
An increase of around \SI{2}{\percent} for the bremsstrahlung cross sections, mainly driven by the radiative corrections, and a slight decrease of the pair production cross section are observed.
Further effects on the propagation are studied using the lepton propagator PROPOSAL.

\subsection{The lepton propagator PROPOSAL}
PROPOSAL \cite{PROPOSAL} is a Monte-Carlo Simulation library to propagate charged leptons.
It is mainly designed to simulate muons with energies above \SI{10}{\giga\electronvolt} traveling large distances through media with high performance and accuracy.
This is needed for large volume detectors like neutrino telescopes or other underground experiments dealing with an atmospheric muon background.
PROPOSAL is used in the IceCube simulation chain for the propagation and decay of muons and taus.

PROPOSAL is a C++ library, originally written in Java (called MMC) \cite{mmc}, but through a pybinding wrapper, it can also be used as a Python library.
The current version \cite{dPROPOSAL}\footnote{The code is available at \texttt{https://github.com/tudo-astroparticlephysics/PROPOSAL}.} includes a complete reconstruction to a modular code structure and polymorphism resulting in a performance improvement of \SI{30}{\percent}.
To reach a high precision in a decent amount of time, PROPOSAL uses interpolation tables during initialization.
Although there is the possibility not to use the interpolations and always integrate over the propagation integrals, this decreases the performance by orders of magnitudes.

One of the main goals of PROPOSAL is the ability to do systematic studies concerning the uncertainties in the muon cross sections.
Therefore multiple cross section parametrizations are available to study their effect on the propagation.
Currently, there are two pair production parametrizations, five bremsstrahlung parametrizations and eight nuclear inelastic interaction parametrizations available.
The production of a muon pair \cite{mpp} and the weak interaction \cite{Sandrock} are optional processes.
Although the muon production doesn't contribute to the average energy loss, the additional muon tracks create a different event signature in a detector and also increase the muon flux for air showers measured on earth.
The weak exchange of a charged current is even less common, but the disappearing high energy muon track in a hadronic shower and an invisible outgoing neutrino creates a unique signature inside the detector.

\subsection{Effects on the range- and energy loss distribution}
\begin{figure}
    \begin{subfigure}{0.46\textwidth}
        \centering
        \includegraphics[width=\textwidth]{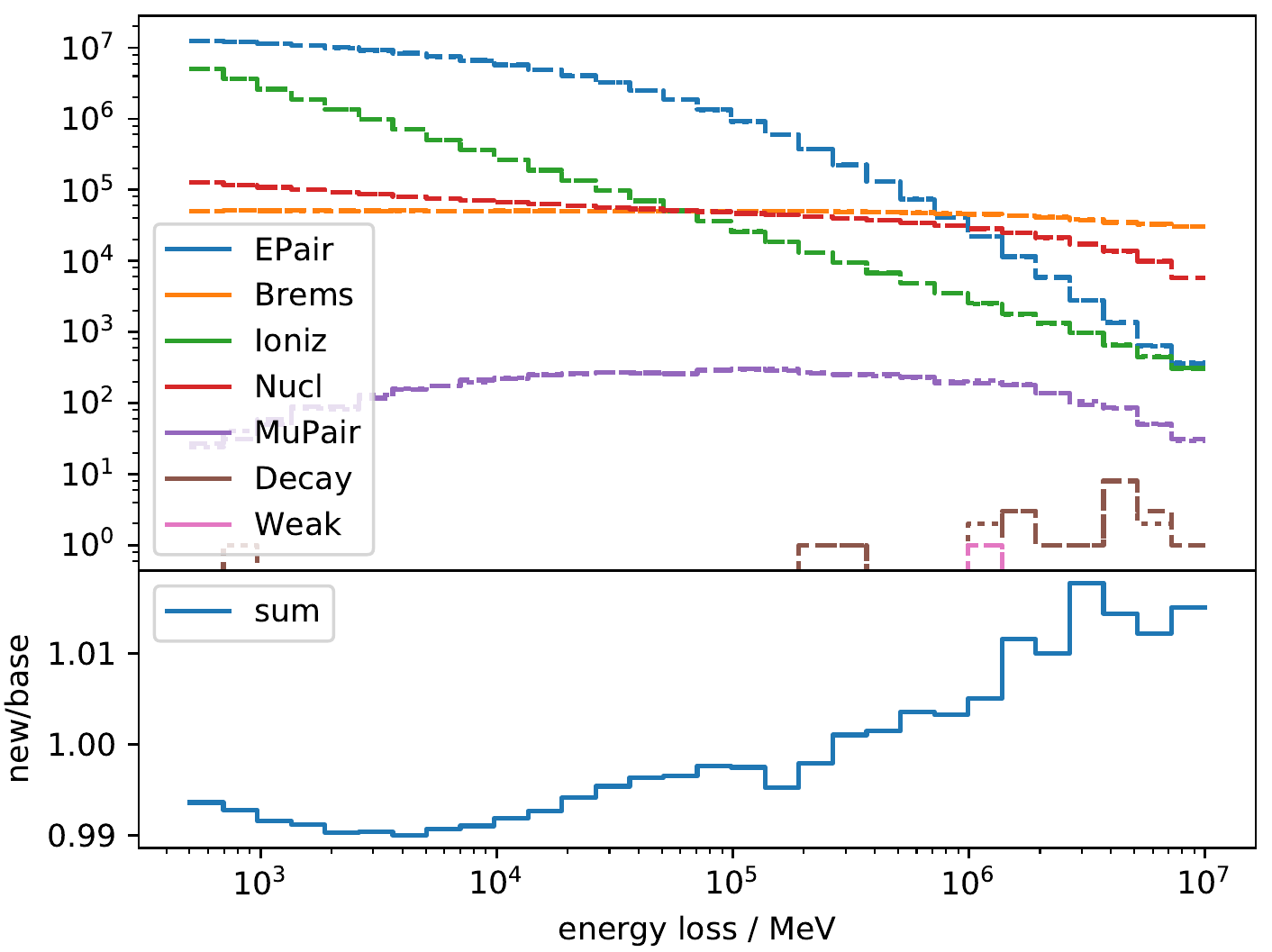}
        \caption{Secondaries distribution of $10^7$ muons propagated \SI{100}{\meter}.}
        \label{fig:sec_dist}
    \end{subfigure}
    \hfill
    \begin{subfigure}{0.46\textwidth}
        \centering
        \includegraphics[width=\textwidth]{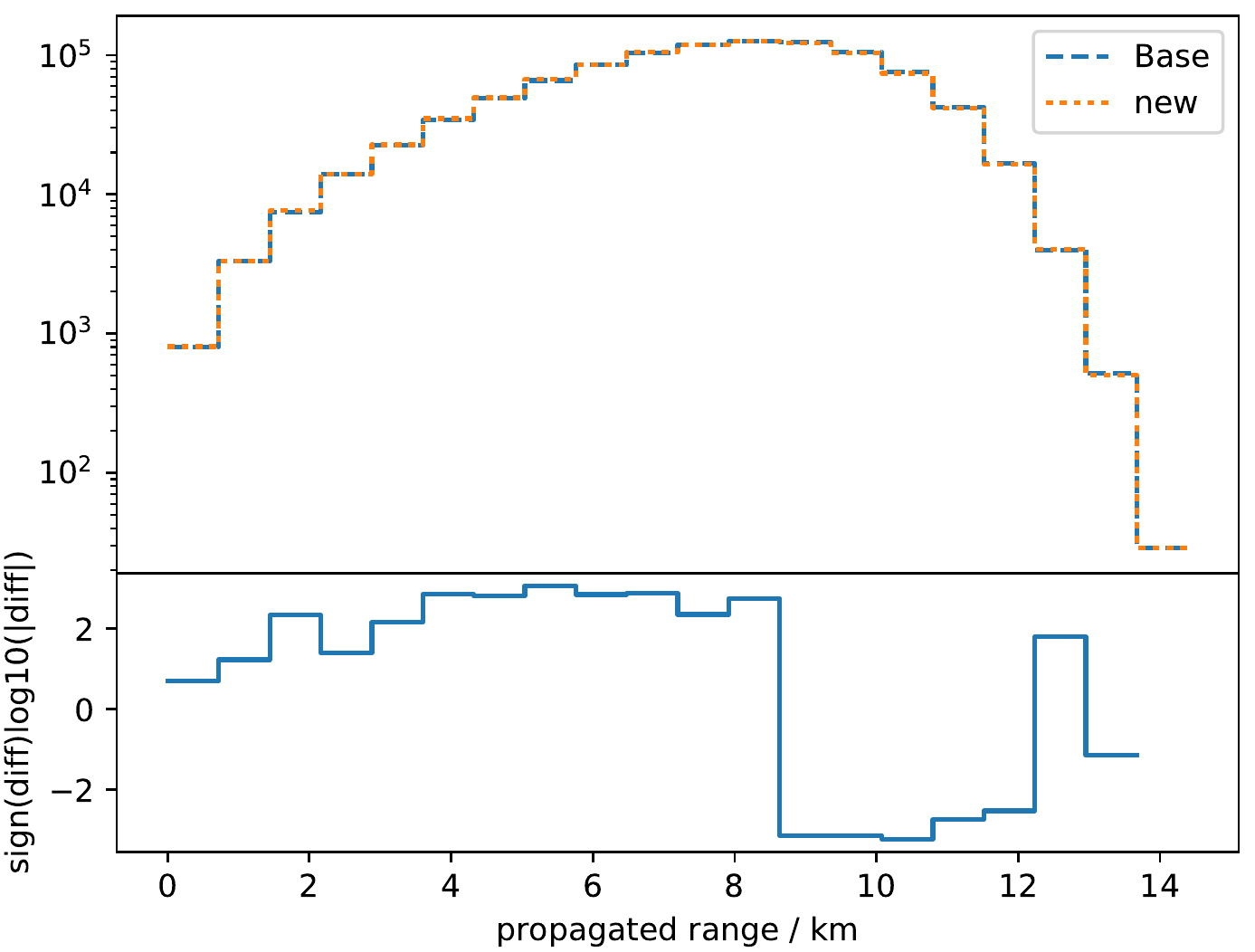}
        \caption{Range distribution of $10^6$ muons propagated until they decay.}
        \label{fig:range_dist}
    \end{subfigure}
    \caption{Effects of the cross sections shown above (dotted) compared to Kelner et al. \cite{KKP,KKP-atomic,KokoulinPetrukhin1,KokoulinPetrukhin2,Kelner-atomic} (dashed) regarding the secondaries \ref{fig:sec_dist} and range \ref{fig:range_dist} distribution when propagating muons with an energy of \SI{10}{\tera\electronvolt} through ice using a minimum energy loss cut of \SI{500}{\mega\electronvolt}.}
    \label{fig:prop_dist_compare}
\end{figure}

The effects of using different cross sections can be best seen in the distribution of the secondary particles, or the energy losses and decay products.
In addition to that, the range distribution should also be affected, while being influenced more indirectly.
A comparison of the bremsstrahlung and pair production cross sections calculated in this work with the widely used cross sections calculated by Kelner et al. \cite{KKP,KKP-atomic,KokoulinPetrukhin1,KokoulinPetrukhin2,Kelner-atomic} on the range and secondaries distribution using PROPOSAL is shown in figure \ref{fig:prop_dist_compare}.

In the secondaries distribution \ref{fig:sec_dist}, an increase at the higher energy losses is observed, which is expected as the new bremsstrahlung cross section is slightly higher.
In addition to that, the pair production cross section is slightly lower, which increases the effect, since the bremsstrahlung dominates the higher energy losses while the pair production dominates the lower energy losses at this energy.
The additional processes of muon pair production and weak interaction are negligible as well as the electrons from decays.

In the range distribution \ref{fig:range_dist}, the higher range bins are decreased, which is explained by the higher cross sections. These shorter propagating muons are then distributed in the lower ranges bins, which then increases.

\section{Conclusion}
New muon cross sections for bremsstrahlung and pair production with an improved description of the screening of an atom is presented showing a decrease of half a percent of the cross section.
In addition radiative corrections for bremsstrahlung showing an increase of \SI{2}{\percent} are presented.
These new cross sections slightly shift the energy loss distribution to more high energy losses and less smaller losses.
Also, the range distribution of the muons tends to shorter ranges.

\section*{Acknowledgments}
We acknowledge funding by the Deutsche Forschungsgemeinschaft (DFG) -- Project number 349068090.

\bibliographystyle{ICRC}
\bibliography{references}

\end{document}